\documentstyle[epsf]{article}
\begin{document}

\begin{flushright}
hep-ph/9512342
\end{flushright}

\begin{center}
{\Large \bf Singular Classical Solutions and Tree Multiparticle Cross
Sections in Scalar Theories}\footnote{To appear in Proc.\ Xth Int.\ Workshop
on high energy physics and quantum field theory, Zvenigorod, 1995.} \\

\vspace{4mm}

F.L.Bezrukov$^{1,2}$,
M.V.Libanov$^{1,2}$,
D.T.Son$^{3}$,
 and S.V.Troitsky$^{2}$

\vspace{4mm}

{\small
$^1$
Physics Department, MSU, 117234
Moscow, Russia

$^2$
 Inst. for Nuclear Research of
the Russian Academy of Sciences, 117312 Moscow, Russia

$^3$
Physics Department, University of Washington, Seattle WA 98195, USA}

\end{center}
%
\begin{abstract}
We consider the features of multiparticle tree cross sections in scalar
theories in the framework of a semiclassical approach. These cross sections
at large multiplicities have exponential form, and the
properties of the exponent in different regimes are discussed.
\end{abstract}

\section{Introduction}

Considerable interest has been attracted in recent years to the issue of
multiparticle production both in perturbative and non-perturbative regimes
in weakly coupled scalar field theory (for a review, see \cite{Volr}).
This problem has been initiated by the qualitative observation
\cite{Corn,Golb} that in the ordinary theory with the action
\begin{equation}
S=\int
d^{d+1}x\left(\frac{1}{2}(\partial_\mu\varphi)^2-\frac{1}{2}\varphi^2-
\frac{\lambda}{4}\varphi^4\right)
\label{1*}
\end{equation}
(hereafter we set the mass of the boson equal to unity) cross
sections of the processes of creation of a large number of bosons by few
initial ones exhibit factorial dependence on the multiplicity of the final
state.  The reason is that the number of tree graphs contributing to the
amplitude to produce $n$ particles grows as $n!$. At $n\sim1/\lambda$ this
factor is sufficient to compensate the suppression due to the small
coupling constant, and the tree level multiparticle cross sections
become large.

Now there is a lot of perturbative results (see, for instance,
\cite{LRST}) which confirm the factorial growth of the tree level
amplitudes, and, what is more important, exhibit the exponential behavior
of multiparticle cross sections
\begin{equation}
\sigma(E,n)\propto\exp\left(\frac{1}{\lambda}F(\lambda n,\varepsilon)\right)
\label{1}
\end{equation}
where $\varepsilon=(E-n)/n$, and $E$ is the energy of initial state.
Though practically all these results have been obtained in the
perturbative regime $\lambda n\ll1$, $\varepsilon\ll1$, one expects that the
exponential behavior survives in the regime $\lambda\to 0$, with $\lambda
n,\,\,\varepsilon$ being fixed. Moreover, there are strong indications
\cite{LST} that the  exponent $F(\lambda n, \varepsilon)$ is universal,
i.e. independent of the few-particle initial state.

The exponential form of $\sigma(E,n)$ implies that for the calculation of
the function $F(\lambda n, \varepsilon)$ there may exist a semiclassical
method like the Landau technique \cite{Landau} for calculating matrix
elements in  quantum mechanics. In fact, some approaches aiming to
generalize the Landau method have been proposed recently
\cite{Son,VolDya,Khleb}.

Let us concentrate on the technique proposed in \cite{Son}. It is based on
singular solutions of some classical boundary value problem. As
the result, the function $F(\lambda n,\varepsilon)$ obtained by
perturbative calculations \cite{LRST} has been reproduced by means of this
way. In the $(d+1)$-space-time it looks like
\begin{equation}
F(\lambda n,\varepsilon)=\lambda n\ln\frac{\lambda n}{16}-\lambda n +
\frac{d\lambda n}{2}\left(\ln\frac{\varepsilon}{\pi d}+1\right)+
\frac{3d-26}{12}\lambda n\varepsilon +B\lambda^2n^2+O(\lambda^3n^3)+
O(\lambda^2n^2\varepsilon)+O(\lambda n\varepsilon^2)
\label{2}
\end{equation}
where $B$ is some known numericalconstant. Moreover, certain new properties
of $F$ have been
obtained in different regimes. The given method is very promising since
it allows one to study $F(\lambda n,\varepsilon)$ by the perturbative
expansion or, in any case, to put this issue in the computer.

In this paper we study  the behavior of
$F(\lambda n,\varepsilon)$ in the regime $\lambda n\ll1,$ with $ \varepsilon$
being fixed, i.e. in fact we will investigate only the energy dependence of
the tree level cross sections. We present first few terms in the low energy
expansion of $F$ in powers of $\varepsilon$ (up to $\lambda
n\varepsilon^2$). A lower bound on the function $F$ at any energies is
 obtained by numerical calculations. This bound seems to be larger than
the estimate found by Voloshin \cite{Volb} from the analysis of the Feynman
diagrams and coincides with the asymptotics at the high energies
\cite{Khleb}. We demonstrate also that in fact the function $F$ lies above
this bound and does not coincide with it.

The paper is organized as follows. In Sect.2 we present a
calculation of the boundary value problem for the tree level
cross sections suggested in \cite{Viet} leading to the result
which coincides with one obtained from the more general approach
\cite{Son}.  In Sect.3 we find the $O(\varepsilon^2)$-correction.
Sect.4 is devoted to the proof that, at least in high energy limit,
spherically symmetric solutions do not provide the true value of $F$. In
Sect.5 some numerical results are presented. In Sect.6 we present our
conclusions. Several technical details are collected in  Appendix.

\section{General formalism}

Now with the aim to introduce our notations and to expose some points of
the calculation we will briefly discuss the method suggested in
\cite{Viet}.  We will examine the cross section of the decay of one
virtual particle at rest with the energy $E$ into $n$ real particles in the
model (\ref{1*}). Let us start from the coherent state representation
(see \cite{Peter} for details) for the matrix element $\langle
\beta|S\varphi|0\rangle$.  In general, it has the form

\[
\langle
\beta|S\varphi|0\rangle=\lim_{T_f\to\infty}\int D\varphi D\varphi_{\! f}
\varphi(E,{\bf k})\exp\{iS+B_f(\beta_k^*,\varphi_f)\}
\]
where
\[
\varphi(E,{\bf k})=\int dtd^dx\varphi(t,{\bf x}){\rm e}^{-iEt+i{\bf kx}},
\,\,\, \varphi_k(t)=\int \frac{d^dx}{(2\pi)^{d/2}}\varphi(t,{\bf
x}){\rm e}^{-i{\bf kx}},
\,\,\,\varphi_f({\bf k})=\varphi_k(T_f)
\]
\[
B_f(\beta_k^*,\varphi_f)=-\frac{1}{2}\int\! d^dk\beta_k^*\beta_{-k}^*
{\rm e}^{2i\omega T_f}-\frac{1}{2}\int\! d^dk\omega\varphi_f({\bf k})
\varphi_f(-{\bf k})+\int\! d^dk\sqrt{2\omega}{\rm e}^{i\omega T_f}
\beta_k^*\varphi_f(-{\bf k}),\,\,\,\omega=\sqrt{1+{\bf k}^2}
\]
To calculate this matrix element at the tree level one should extremize
$(iS+B_f)$ over $\varphi$ and $\varphi_f$. This extremization yields the
classical field equation with the boundary conditions
\begin{equation}
\partial_\mu^2\varphi+\varphi+\lambda\varphi^3=0,\,\,\,\,
\varphi_k(t\to+\infty)=\frac{\beta_k^*}{\sqrt{2\omega}}{\rm e}^{i\omega t}
,\,\,\,\, \varphi_k(t\to-\infty)\sim a_k{\rm e}^{i\omega t}
\label{4}
\end{equation}
The solution to this equation, $\varphi_c(\beta^*,x)$ has only positive
frequency parts due to the boundary conditions, and hence the action $S$ and
the
boundary term $B_f$ are reduced to zero. Thus, the tree level matrix
element is

\[
F(\beta^*)\equiv\langle \{\beta\}|S\varphi|0\rangle_{{\rm tree}}=\int dtd^dx
\varphi_c(\beta^*,t,{\bf x}){\rm e}^{-iEt+i{\bf kx}}
\]
and the amplitude $1\to n$ is given by
\[
A_n=\frac{\partial^n}{\partial\{\beta^*\}^n}F(\beta^*)\Big|_{\{\beta^*\}=0}
\]
This result  coincides with one obtained in \cite{Brown} from LSZ-reduction
formula.

According to the coherent state formalism the $n$-particle cross section
can be easily found,
\begin{equation}
\sigma_{1\to n}^{{\rm tree}}=\oint\frac{d\xi}{\xi}\int D\beta D\beta^*
 dxdx' \varphi_c(\beta^*,x) \varphi_c^*(\beta,x')\exp\left\{-iE(t-t')+i
{\bf k(x-x')}-\frac{1}{\xi}\int d^dk\beta^*\beta - n\ln\xi\right\}
\label{5}
\end{equation}
where the integral over $\xi$ corresponds to the $n$-th derivative due to
Cauchy theorem. In order to estimate the integral (\ref{5}) the saddle point
technique can be applied. To perform it correctly, however, one should look
more carefully at $\varphi_c$ and get rid of  zero modes in $\varphi_c$
corresponding to time shifts. One writes
\[
\beta^*_k=b^*_k{\rm e}^{i\omega\eta}\,\,\Rightarrow\,\,
\varphi_c(\beta^*,t)=\varphi(b^*,t+\eta)
\]
where $\eta$ is the collective coordinate, and  $b_k$ are new integration
variables
which obey a constraint to be spacified below.
(Of course, there are zero modes corresponding to the space shifts, but
we do not write them explicitly  since taking them into account leads only
to spacial momentum conservation). So,  Eq.(\ref{5}) becomes
\[
\sigma_{1\to n}^{{\rm tree}}=\oint\frac{d\xi}{\xi}\int Db Db^*
 d\eta d\eta'\varphi(b^*,E,{\bf k}) \varphi^*(b,E,{\bf
 k})\exp\left\{-iE(\eta-\eta')-\frac{1}{\xi}\int
 d^dkb^*b{\rm e}^{i\omega(\eta'-\eta)} - n\ln\xi\right\}
\]
Now, if one imposes a constraint that $\varphi$ does not contain exponents
and, so, may be considered as a pre-exponential factor, then this integral
can be calculated by making use of the saddle point technique. The exact
form of this constraint, which may be viewed as a constraint on $b_k^*$,
 will be determined in what follows. Therefore,
 the tree cross section is
\begin{equation}
\sigma(E,n)^{{\rm tree}}\propto{\rm e}^{W_{{\rm extr}}}
\label{6*}
\end{equation}
where $W_{{\rm extr}}$ is the saddle point value of the functional
\begin{equation}
W=ET-n\theta-\int d^dkb_k^*b_k{\rm e}^{\omega T-\theta}
\label{6}
\end{equation}
with respect to
$T=i(\eta'-\eta)$, $\theta=\ln\xi$, $b_k$, and $b_k^*$
under an  additional constraint which is yet to be defined.

In general, one expects that these classical solutions, being considered in
Euclidean space-time, have singularities  on some surface $\tau=\tau({\bf x})$.
The leading behavior of the integral
\[ \varphi(E,{\bf k})=\int d^{d+1}x\varphi(x){\rm e}^{-iEt+i{\bf kx}}
\]
is determined by the nearest
to the origin singularity of $\varphi(x)$ in the region $-\tau={\rm
Im}\,\,t<0$. If this singularity is located at $\tau=\tau^*$, then
\[
\varphi(E,{\bf k})\propto{\rm e}^{-E\tau^*}
\]
So, the constraint that $\varphi$ may be considered as a pre-exponential
factor requires $\tau^*=0$. Since we are working in the center-of-mass
frame, the result should depend on $|{\bf k}|$, but not on its direction.
So, in ${\bf x}$-space we require $O(3)$ spherical symmetry and finally the
constraint is that $\varphi(x)$ has the singularity at $\tau=0$, ${\bf x}=0$.

Therefore, the problem of finding the cross section at any value of $E$ and
$n$ can be formulated in  Euclidean space-time and consists of the
following steps,
\begin{itemize}
\item{} One selects from all solutions $\varphi(\tau,{\bf x})$ to the
Euclidean field equation,  $O(3)$-symmetrical ones which are
singular at $\tau={\bf x}=0$ and have the following asymptotics at
$\tau\to\infty$,
\begin{equation}
\int\frac{d^dx}{(2\pi)^{d/2}}\varphi(\tau,{\bf x}){\rm e}^{-i{\bf kx}}=
\frac{b_k^*}{\sqrt{2\omega}}{\rm e}^{-\omega\tau}
\label{7}
\end{equation}
{}From them one finds Fourier components $b_k$ and then  determines $W$ from
Eq.(\ref{6}).
\item{} One should extremize $W$ over all  $b_k$, $b_k^*$
(that is over different singular surfaces), $T$, and $\theta$. The tree
level cross section is then given by the formula (\ref{6*}).
\end{itemize}

Now let us see that the extremum of $W$ at the fixed energy $\varepsilon$
and $n$ is in fact  its maximum. Indeed, let us consider the functional
\[
\int d^dkb_kb_k^*{\rm e}^{\omega T}
\]
This functional is bounded below, so it has a minimum at some value of
$b_k$ (at this point $W$ as a function of $b_k$ has its maximum),
\[
\int d^dkb_kb_k^*{\rm e}^{\omega T}\Big|_{{\rm min}}=C(T)>{\rm const}
\]
If one fixes some particular surface of singularities $\Sigma$, than the
obtained value of the functional
\[
\int d^dkb_kb_k^*{\rm e}^{\omega T}\Big|_{\Sigma_{{\rm fixed}}}=C_\Sigma(T)
\ge C(T)
\]
for all values of $T$. Taking a saddle point value of $\theta$, one gets
$\theta=-\ln n+\ln C(T)$ and
\[
 W(T)=n\ln n-n+ET-n\ln  C(T)\,\,\,\,\mbox{has an extremum at
} T_1,
\]
\[
 W(T)_\Sigma=n\ln n-n+ET-n\ln  C_\Sigma(T)\,\,\,\,\mbox{has  an
extremum at } T_2
\]
Comparing $W(T_1)$ and $W(T_2)$ the following chains
can be obtained
\[
 W(T_1)\ge W(T_2)\ge W_\Sigma(T_2)\,\,\,\,\mbox{ if $T_1$
provides a maximum of $W$}
\]
\[
 W_\Sigma(T_2)\le W_\Sigma(T_1)\le
W(T_1)\,\,\,\,\mbox{ if $T_2$ provides a minimum of $W_\Sigma$}
\]
 Therefore, if one
fixes some class of surfaces of singularities  and energy $\varepsilon$ or,
in other words, finds some particular singular solution, than one can get a
lower bound for $W(E,n)$.

Due to the nonlinearity of the field equations it is not possible to obtain
an analytical solution to the boundary value problem formulated above.
One can find, however, analytical solutions in several regimes.

\section{Low energy expansion of $F(\lambda n,\varepsilon)$}

In this section we calculate $F(\lambda n,\varepsilon)$ to the accuracy of
$O(\varepsilon^3)$. The typical momentum of final particles at small
$\varepsilon$ is much smaller than the mass, so one can expect that the
solution to the field equation $\varphi_c$ is a slowly varying function of
$x$.  The $x$-independent solution having singularity at $\tau=0$ and decaying
at $\tau\to\infty$ can be found exactly \cite{Brown},
\[
\varphi_0(\tau)=\sqrt{\frac{2}{\lambda}}\frac{1}{\sinh\tau}
\]
It corresponds, however, to the case of all final particles being at rest.
Let us modify this solution by imposing by hand dependence on ${\bf x}$.
One writes
\[
\varphi_0(\tau,{\bf x})=\sqrt{\frac{2}{\lambda}}\frac{1}{\sinh(\tau-
\tau_0({\bf x}))}
\]
This function has the surface of singularities $\tau=\tau_0({\bf x})$.
In the end we will extremize the functional $W$, Eq.(\ref{6}), with respect to
$\tau_0({\bf x})$. According to the general formalism described above,
$\tau_0(0)$ must vanish, and
due to spherical symmetry $\tau_0({\bf x})$ depends only on $|{\bf x}|$.
Starting from this function (which satisfies the field equation to the
accuracy of $O((\partial_{\bf x}\tau_0)^2)$) one can expand $\varphi$ in the
following way,
\begin{equation}
\varphi=\varphi_0+\varphi_1+\varphi_2+\dots
\label{9}
\end{equation}
where $\varphi_1,\,\,\,\varphi_2$ are of order of $(\partial_{\bf
x}\tau_0)^2 ,\,\,\,(\partial_{\bf x}\tau_0)^4$ respectively.
  The explicit form of $\varphi_1$, $\varphi_2$ can be found in
Appendix.

In order to obtain $b_k^*$
corresponding to the solution, let us find the asymptotics of
$\varphi$ at
$\tau\to\infty$ (see Eq.(\ref{2*'}), $z=\tau-\tau_0({\bf x})$),
\[
\varphi^{{\rm as}}=\sqrt{\frac{2}{\lambda}}{\rm e}^{-z}\left\{
2-\frac{5}{6}\partial^2\tau_0+(\partial^2\tau_0+(\partial\tau_0)^2)z+
\partial^4\tau_0\left(\frac{z^2}{4}-\frac{z}{6}-\frac{1}{8}\right)+
\right.
\partial^2(\partial\tau_0)^2\left(\frac{z^2}{4}+\frac{z}{4}
 -\frac{3+\pi^2}{24} \right)
\]
\[
+\partial\tau_0\partial^3\tau_0\left(\frac{z^2}{2}-\frac{4z}{3}+
\frac{3+\pi^2}{12} \right)+
\partial\tau_0\partial(\partial\tau_0)^2
\left(\frac{z^2}{2}-\frac{z}{2}+\frac{5}{12}\right)+
(\partial^2\tau_0)^2\left(\frac{z^2}{4}-\frac{2z}{3}+\frac{\pi^2-7}{12}
\right)
\]
\begin{equation}
\left.
+(\partial\tau_0)^2\partial^2\tau_0\left(\frac{z^2}{2}-\frac{17z}{12}
+ \frac{5}{12}\right)+
(\partial\tau_0)^4\left(\frac{z^2}{4}-\frac{3z}{4}
\right)\right\}
\label{3'}
\end{equation}
This expression is of  course the solution to the accuracy of
$O((\partial\tau_0)^6)$ to the linearized field equation which has
the following general solution,
\begin{equation}
\varphi(x,\tau)=\int\frac{d^dk}{(2\pi)^{d/2}}\frac{b_k^*}{\sqrt{2\omega}}
{\rm e}^{i{\bf kx}-\omega\tau},\,\,\,\,\,\,\,\omega=\sqrt{1+k^2}
\label{*}
\end{equation}

Let us introduce the notation
\begin{equation}
b(x)=\sqrt{\frac{\lambda}{2}}\int\frac{d^dk}{(2\pi)^{d/2}}\frac{b_k^*}
{\sqrt{2\omega}} {\rm e}^{i{\bf kx}}
\label{3**'}
\end{equation}
Thus, one  gets, for any function of the form (\ref{*}), the following
relation,
\begin{equation}
\varphi({\bf x},\tau)=\int\frac{d^dk}{(2\pi)^{d/2}}\frac{b_k^*}
{\sqrt{2\omega}} {\rm e}^{i{\bf kx}-\tau}(1-\frac{\tau }{2}k^2+\frac{\tau+
\tau^2}{8}k^4)=
{\rm e}^{-\tau}\sqrt{\frac{2}{\lambda}}(b+\frac{\tau}{2}\partial^2b+
\frac{\tau+\tau^2}{8}\partial^4b)
\label{3+'}
\end{equation}

Assuming that
\begin{equation}
b=2{\rm e}^{\tau_0}+b_1+b_2
\label{4+'}
\end{equation}
where $b_1\sim(\partial\tau_0)^2$, $b_2\sim(\partial\tau_0)^4$,
one can find from (\ref{3+'}) that to the accuracy of
$O((\partial\tau_0)^6)$,
\begin{equation}
\varphi=\sqrt{\frac{2}{\lambda}}{\rm e}^{-\tau}(2{\rm e}^{\tau_0}
+\tau\partial^2{\rm e}^{\tau_0}+b_1+\frac{\tau+\tau^2}{4}
\partial^4{\rm e}^{\tau_0}+\frac{\tau}{2}\partial^2b_1+b_2)
\label{4'}
\end{equation}
Comparing the terms of the same order in $\partial\tau_0$ in Eqs.(\ref{4'})
and (\ref{3'}) one finds for $b_1(0)$ and $b_2(0)$,
\begin{equation}
b_1(0)=-\frac{5}{6}\partial^2\tau_0
,\,\,\,\,b_2(0)=-\frac{\partial^4\tau_0}{8}-\frac{3+\pi^2}{12}
(\partial_{ij}^2\tau_0)^2+\frac{\pi^2-7}{12}(\partial^2\tau_0)^2
\label{4*'}
\end{equation}
To get these expressions we took into account that $\tau_0(0)=0$ and due to
the $O(3)$ spherical symmetry, $\partial_i\tau_0(x)|_{x=0}=0$.
{}From Eqs.(\ref{4*'}) and  (\ref{4+'}) one finds the
following system of the equations for $b$ and its derivatives at ${\bf
x}=0$,
\[
 b=2-\frac{5}{6}\partial^2\tau_0-\frac{1}{8}\partial^4\tau_0-
\frac{3+\pi^2}{12}(\partial_{ij}^2\tau_0)^2+\frac{\pi^2-7}{2}
(\partial^2\tau_0)^2\,\,;
\]
\begin{equation}
\partial_{ij}^2b=2\partial_{ij}^2\tau_0-\frac{5}{6}\partial_{ij}^2\tau_0
\partial^2\tau_0-\frac{11}{6}\partial_{ij}^2\partial^2\tau_0,
\,\,\,\,
\partial^4b=2\partial^2\tau_0+4(\partial_{ij}^2\tau_0)^2+
2(\partial^2\tau_0)^2
\label{5'}
\end{equation}
{}From this system one can easily find all the derivatives of $\tau_0$ with
respect to ${\bf x}$ at ${\bf x}=0$ via the derivatives of $b$,
\begin{equation}
\partial^2\tau_0=\frac{1}{2}\partial^2b+\frac{5}{24}\partial^4b-\frac{5}{24}
(\partial_{ij}^2b)^2+\frac{1}{8}(\partial^2b)^2
,\,\,\,\,
(\partial_{ij}^2\tau_0)^2=\frac{1}{4}(\partial_{ij}^2b)^2,\,\,\,\,
\partial^4\tau_0=\frac{1}{2}\partial^4b-\frac{1}{2}
(\partial_{ij}^2b)^2-\frac{1}{4}
(\partial^2b)^2
\label{5*'}
\end{equation}
Substituting these solutions into the first equation of (\ref{5'}) one gets
the constraint on $b$,
\begin{equation}
2=b+\frac{5}{12}\partial^2b+\frac{17}{72}\partial^4b+\frac{3\pi^2-25}{144}
(\partial_{ij}^2b)^2+\frac{21-2\pi^2}{96}(\partial^2b)^2
\label{5**}
\end{equation}
{}From  (\ref{3**'}) and (\ref{5**}) one obtains
\[
\sqrt{\frac{16}{\lambda}}=
\sqrt{\frac{
\lambda}{4}}\frac{3\pi^2-25}{144}
\left(\int\frac{d^dk}{(2\pi)^{d/2}}\frac{k_ik_jb_k^*}{\sqrt{\omega}}\right)^2+
\sqrt{\frac{\lambda}{4}}\frac{21-2\pi^2}{96}
\left(\int\frac{d^dk}{(2\pi)^{d/2}}\frac{k^2b_k^*}{\sqrt{\omega}}\right)^2
\]
\begin{equation}
+\int\frac{d^dk}{(2\pi)^{d/2}}\frac{b_k^*}
{\sqrt{\omega}}\left(1-\frac{5}{12}k^2+\frac{17}{72}k^4\right)
\equiv{\cal{F}}(b^*_k)
\label{5***'}
\end{equation}
Let us now to extremize (\ref{6}) taking into account the
constraint (\ref{5***'}). Making use of the Lagrange multiplier method,
we rewrite (\ref{6}) in the following form
\begin{equation}
W=ET-n\theta-\int d^dkb_kb_k^*{\rm e}^{\omega T-\theta}+
A\left({\cal{F}}(b_k^*)-\sqrt{\frac{16}{\lambda}}\right)
\label{6'}
\end{equation}
where the constant $A$ should be determined. Now $W$ can be variated with
respect to $b_k^*$ without any restrictions. Performing this procedure one
gets
\begin{equation}
b_k=\frac{A{\rm e}^{-\omega T+\theta}}{(2\pi)^{d/2}\sqrt{\omega}}
\left(1-\frac{5} {12}k^2+\frac{17}{72}k^4+2\sqrt{\frac{\lambda}{4}}
\left(\frac{3\pi^2-25}{144} k_ik_jC_{ij}+
\frac{21-2\pi^2}{96}k^2{\rm Tr}C_{ij}\right)\right)
\label{6*'}
\end{equation}
where
\begin{equation}
C_{ij}=\int\frac{d^dk}{(2\pi)^{d/2}}\frac{k_ik_j}{\sqrt{\omega}}b_k^*
\label{6**'}
\end{equation}
Assuming $b_k=b_k^*$, from (\ref{6*'}) and (\ref{6**'}) one
obtains the equation for $C_{ij}$,
\begin{equation}
 C_{ij}=A\int\frac{d^dk}{(2\pi)^d}\frac{k_ik_j}{\omega}
 {\rm e}^{-\omega T+\theta}
\left(1-\frac{5}{12}k^2+\frac{17}{72}k^4+2\sqrt{\frac{\lambda}{4}}
\left(\frac{3\pi^2-25}{144}k_nk_mC_{nm}+
 \frac{21-2\pi^2}{96}k^2{\rm Tr}C_{nm}\right)\right)
\label{6+'}
\end{equation}
Let us now concentrate on the equation (\ref{6+'}). First of all we
consider the case $i\ne j$. Due to the fixed parity of the integrand,
\[
C_{ij}=\sqrt{\frac{\lambda}{4}}\frac{A}{(2\pi)^d}\frac{3\pi^2-25}{36}C_{ij}
\int\frac{d^dk}{\omega}(k_i)^2(k_j)^2{\rm e}^{-\omega T+\theta}
\]
(no summations over repeated indices).
{}From the last equation one concludes that $C_{ij}=\delta_{ij}{\rm Tr}
C_{nm}/d \equiv\delta_{ij}C/d$. Together with Eq.(\ref{6+'}), this leads to
\begin{equation}
C= \frac{A}{(2\pi T)^{d/2}}{\rm e}^{\theta-T}\left(\frac{d}{T}+
O(\frac{1}{T^2})\right)\,\,\,\,\,
\label{7*'}
\end{equation}
$T$ is assumed to be large at small $\varepsilon$ since, in fact,
$T\sim W_{{\rm extr}}/E$.
The constant $A$ can then be found from the constraint (\ref{5***'})
by making use of Eqs.(\ref{6*'}) and (\ref{7*'}):
\begin{equation}
A=\sqrt{\frac{16}{\lambda}}\frac{{\rm e}^{T-\theta}}{(2\pi T)^{d/2}}
\left(1-\frac{d(3d-26)}{24T}+
\frac{d}{T^2}\left(\frac{9d^3-156d^2+884d-1632}{1152}-3d\delta\right)\right)
\label{8'}
\end{equation}
where
\[
\delta\equiv\frac{3\pi^2-25}{72d} +\frac{21-2\pi^2}{48}
\]
So, one finally obtains the  expression for $b_k$,
\[
b_k=\sqrt{\frac{16}{\lambda}}\frac{{\rm e}^{T-\omega T}T^{d/2}}{\omega}
\left(1-\frac{d(3d-26)}{24T}+
\frac{d}{T^2}\left(\frac{9d^3-156d^2+884d-1632}{1152}-3d\delta\right)
\right)
\]
\begin{equation}
\times\left(1+\left(\frac{2d\delta}{T}-\frac{5}{12}\right)k^2+
\frac{17}{72}k^4\right)
\label{8*'}
\end{equation}
The variation of the functional (\ref{6'})  with respect to $\theta$
yields to
\begin{equation}
 n=\frac{A^2{\rm e}^{\theta-T}}{(2\pi
T)^{d/2}}\left(1+\frac{d(3d-26)}{24T}+
\frac{d}{T^2}\left(\frac{(d+2)(3d^2-58d+272)}{384}+4d\delta\right)\right)
\label{8**'}
\end{equation}
{}From (\ref{8**'}) $\theta$ can be found as a function of $T$ and $n$.
The variation of (\ref{6'}) with respect to $T$ gives the equation for $T$,
\begin{equation}
\varepsilon=\frac{d}{2T}\left(1+\frac{3d-26}{12T}-\frac{13d-102}{18T^2}+
\frac{8d\delta}{T^2}\right)\,\,\,\,\Rightarrow
T=\frac{d}{2\varepsilon}+\frac{3d-26}{12}+\varepsilon\left(16\delta-
\frac{9d^2-52d-140}{72d}\right)
\label{8***'}
\end{equation}
$T$ is seen to be indeed large at small $\varepsilon$.

Collecting (\ref{8'}), (\ref{8**'}), and (\ref{8***'})
one finally gets for the function $W$
\begin{equation}
W=n\left(\ln\frac{\lambda n}{16}-1+\frac{d}{2}\left(\ln\frac{\varepsilon}
{d\pi}+1\right)+\varepsilon\frac{(3d-26)}{12}-\frac{\varepsilon^2}{144d}
(9d^2-556d+260+48\pi^2(d-1))\right)
\label{9'}
\end{equation}
The first three terms in this expression coincide with those obtained in
\cite{Son} and the last term is an $\varepsilon^2$ correction to $W$.
Particularly, at
$d=3$, $W$ has the  form
\begin{equation}
W_{d=3}=n\left(\ln\frac{\lambda n}{16}-1+\frac{3}{2}
\left(\ln\frac{\varepsilon}{3\pi}
+1\right)-\varepsilon\frac{17}{12}+
\frac{\varepsilon^2}{432}(1327-96\pi^2)\right)
\label{9*'}
\end{equation}

Finally, the last thing which is interesting to be performed is the
calculation of $\tau_0(x)$ which corresponds to the obtained $b_k$
(\ref{8*'}). To do it, one can find all derivatives of $\tau_0(x)$ at $x=0$
(see (\ref{5*'})) and after that restore $\tau_0(x)$ as the Taylor's series
in powers of $|{\bf x}|$.

So, using the definition of $b$ (\ref{3**'}), expression for $b_k$
(\ref{8*'}), and taking into account that due to spherical symmetry
$\partial_{ij}^2=\frac{\delta_{ij}}{d}\partial^2$ one finds
\begin{equation}
\partial^2b\big|_{x=0}=-\frac{2d}{T}-\frac{d(11d-10)}{6T^2}+
O\left(\frac{1}{T^3}\right),\,\,\,\,\,
\partial^4b\big|_{x=0}=\frac{2d(d+2)}{T^2}+O\left(\frac{1}{T^3}\right)
\label{10'}
\end{equation}
All  derivatives of $\tau_0$ can be easily found from  Eqs.
(\ref{5*'}) and (\ref{10'}),
\[
\partial^2\tau_0\big|_{x=0}=-\frac{d}{T}+\frac{5}{6}\frac{d}{T^2},\,\,\,\,\,
\partial_{ij}^2\tau_0\big|_{x=0}=\delta_{ij}\left(-\frac{1}{T}+\frac{5}{6T^2}
\right),\,\,\,\,\,
\partial^4\tau_0\big|_{x=0}=\partial_{ijkm}^4\tau_0\big|_{x=0}=
\partial^2\partial_{ij}^2\tau_0\big|_{x=0}= 0
\]
Thus, the singularity surface which corresponds to the cross section up to
$O(\varepsilon^2)$ is a paraboloid,
\begin{equation}
\tau_0(x)=x^2\left(-\frac{1}{2T}+\frac{5}{12}\frac{1}{T^2}\right)=
-\frac{x^2\varepsilon}{d}\left(1-\varepsilon\frac{3d-16}{6d}\right)
\label{10*'}
\end{equation}
The Eq.(\ref{10*'}) is valid for values of {\bf x} being smaller
than the typical inverse momentum of the final particles,
$\varepsilon^{-1/2}$. It is worth noting that the term of order $(x^4
\varepsilon^2)$ vanishes.
These results are useful to compare with numerical calculations (see
Sect.5).

\section{Lower bound for $F(\lambda n, \varepsilon)$ in ultrarelativistic
regime}

Let us now turn to the opposite limit, $\varepsilon \gg 1$. In this case
one can neglect the mass term in the field equation and consider the massless
theory.  Since Euclidean massless $\phi^4$ theory is conformally invariant,
several solutions can be found analytically. The simplest one is $O(4)$
symmetric Lipatov's instanton \cite{lipatov}. The corresponding solution
which is singular at
$\tau={\bf x}=0$
can be easily constructed \cite{Khleb} by
making the size of the instanton pure imaginary and shifting its center:
\begin{equation}
\varphi_0=
\sqrt{8\over\lambda}
{\rho\over {\bf x}^2+(\tau+\rho)^2-\rho^2}
\label{Khl}
\end{equation}
The variational procedure on $T$, $\theta$ and only collective coordinate
$\rho$ (clearly, no collective coordinates describing the shape of the
singularity surface appears) could be made and results in (see
\cite{Khleb})
\begin{equation}
W=n\ln{\lambda n\over 16}-\lambda n+n\ln{2\over\pi^2}
\label{WKhl}
\end{equation}
However, the variation on the shape of singularity surface is still required,
unless the result (\ref{WKhl}) provides only lower bound for the true
exponent.

To see that these $O(4)$ symmetric calculations in fact do not provide the
extremum over the shape of singularities, let us consider the small
perturbations about the solution (\ref{Khl}). Since the solution should be
$O(3)$ symmetric with respect to ${\bf x}$ rotations, it is sufficient to
consider the approximate solutions which slightly vary the singularity
surface so that the dependence of $\psi$, an angle between the $\tau$ axis
and the radius vector, appears.  The solution should have the form
\[
\varphi={2\sqrt{\lambda/2}\rho +\alpha f(r)C^{(1)}_m(\cos\psi)
\over r^2-\rho^2 (1+\alpha C^{(1)}_m(\cos\psi))^2}=
\varphi_0+\sum_m \alpha_m C^{(1)}_m(\cos\psi) \tilde{\varphi}_m(r)+ O(\alpha^2)
\]
In this equation, $C^{(1)}_m(\cos\psi)$ are Gegenbauer  polynomials,
$\alpha_m$ are small parameters, and the linear fluctuations
 $\tilde{\varphi}_m$ should satisfy the following equation,
\[
\left[
\partial^2_r+{3\over r}\partial_r-{m(m+2)\over r^2}-
{24\rho^2\over (r^2-\rho^2)^2}
\right]
\tilde{\varphi}_m(r)=0
\]
To solve this equation, one can exploit the conformal invariance of the
model and use the stereographic projection onto the sphere $S^4$ (it is a
convenient way to solve the equations in the conformal models, we use here
a slightly modified method of \cite{kubyshin}). Skipping the details, the
solution is
\[
\tilde{\varphi}_m(r)={1\over\sqrt{\lambda}}
{r^{2-m}\rho^{m+1}\over (r^2-\rho^2)^2}
\left(1-{2(m-1)\over m+2}\,{\rho^2\over r^2} +{(m-1)m\over (m+2)(m+3)}\,
{\rho^4\over r^4}\right)
\]
Calculating the asymptotics of this solution, one obtains the following
expression for $b_k$:
\[
b_k=b_k^0+\sum_m\alpha_m\sqrt{\pi\over k\lambda}{\rm e}^{-k\rho}
\left(\rho^{m+1}{k^m\over m!}-3\rho^2 k {(m+1)\over (m+2)(m+3)}\right)
\]
(here $b_k^0$ corresponds to the solution (\ref{Khl})). Then
the exponent after the variational procedure on $T$ and $\theta$ is
modified by
$$
\Delta W=\sum_m\alpha_m {(m+1)\over\rho}\left(2^{-m-2}-
{3\over 2(m+2)(m+3)}\right)
$$
For $m=0$ or $1$ (which corresponds to the dilatations and translations of
$\varphi_0$), $\Delta W=O(\alpha^2)$. However, for all other values of $m$
the perturbation in the exponent is proportional to $\alpha_m$, so that the
functional derivative of the exponent calculated at $O(4)$-symmetric
solution with respect to collective coordinates describing the shape of the
singularity surface  is non-zero. So, the result (\ref{WKhl}) is indeed
only lower bound on the tree cross sections in high-energy limit.  One
could expect that for lower energies (massive theory) this property will be
conserved.

\section{Lower bound of $F(\lambda n,\varepsilon)$ at arbitrary
$\varepsilon$}

So, to obtain a lower bound for $\sigma_{\rm tree}(E,n)$ at given $E$ and
$n$ one could consider $O(4)$ symmetric field configurations \cite{BLT}.
The singularity surfaces are then three-spheres in Euclidean space, and
the only variational parameter is the radius of a sphere, $\rho$.  Since
the singularity surface should touch the origin, the sphere
should be centered at
$$x_0=(-\rho,{\bf 0})$$
i.e., the $O(4)$ symmetric solution has the general form
$$
\varphi=\varphi(r),\,\,\,\,\,\,
r=\sqrt{{\bf x}^2+(\tau+\rho)^2}
$$
At large $r$, the solution tends to the exponentially falling solution to
the free field equation,
$$
\varphi\sim{K_1(r)\over r}\sim{{\rm e}^{-r}\over r^{3/2}}
$$
i.e., at $\tau\to\infty$ one has
$$
\varphi=A(\rho){\exp(-\sqrt{{\bf x}^2+(\tau+\rho)^2}) \over
({\bf x}^2+(\tau+\rho)^2)^{3/4} }
$$
where the coefficient function $A(\rho)$ is to be determined by solving
the field equations
under the condition that it has a
singularity at $r=\rho$. From
this asymptotics one finds
$$
b_k=2A(\rho)\frac{{\rm e}^{-\omega\rho}}{\sqrt{2\omega} }
$$
The ''action'' (\ref{6}) is then expressed through $A(\rho)$, so
$W$  reads
\begin{equation}
W=ET-n\theta-8\pi A^2(\rho){\rm e}^{-\theta} {K_1(2\rho-T) \over 2\rho-T}
\label{W}
\end{equation}

Now we extremize Eq.(\ref{W}) with respect to $\rho$, $T$ and $\theta$.
This leads to the following equations which determine the saddle
point values of these three parameters,
\begin{equation}
{E\over n}={A'(\rho)\over A(\rho)}= {K_2(2\rho-T) \over K_1(2\rho-T)}
\label{7''}
\end{equation}

\begin{equation}
\theta=\ln \left\{
{8\pi A^2(\rho) \over n} \, {K_1(2\rho-T) \over 2\rho-T} \right\}
\label{5***}
\end{equation}

  So, we look for the classical solutions which are singular at the
spheres $r^2=\rho^2$ and from their asymptotics obtain
$A(\rho)$, then express saddle point values of $\rho$, $T$ and $\theta$
through $E$ and $n$ (by making use of the Eqs. (\ref{7''}), (\ref{5***}))
and finally obtain the estimate for the exponent for the
tree cross section (see Eq.(\ref{6*}))
$$
W_{\rm tree}=ET-n\theta-n
$$

\begin{figure}[tb]
  \begin{center}
    \makebox{\epsfbox{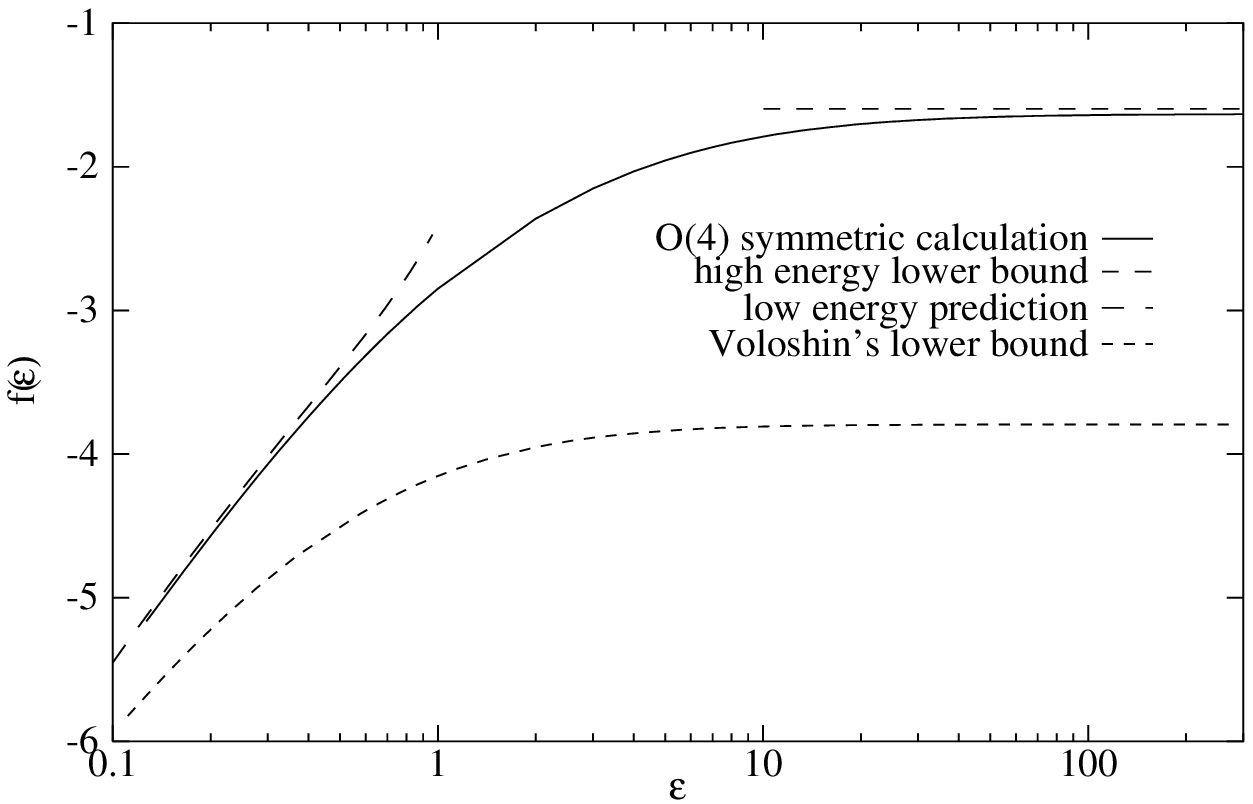}}

    \centerline{Fig.~1}
  \end{center}
\end{figure}

It is straightforward to perform this calculation numerically for all
$\varepsilon$.  The exponent for the cross section has the form
\[
F(\lambda n,\varepsilon)=\lambda n\ln\frac{\lambda n}{16}-\lambda n +
\lambda nf(\varepsilon)+O(\lambda^2n^2)
\]
with the function $f(\varepsilon)$ plotted in Fig.1.  At low
energies our result matches the perturbative results, Eq.(\ref{9*'}).  The
fact that our variational approach leads to the exact results for $W_{\rm
tree}$ at small $\varepsilon$ can be understood as follows. At small
$\varepsilon$, the curvature of the singularity surface is always large,
and only this curvature is relevant for the evaluation of $f(\varepsilon)$
\cite{Son}. In other words, the surface of singularities has the form
(see Eq.(\ref{10*'})),
$$
\tau({\bf x})=a{\bf x}^2+O({\bf x}^4)+\cdots
$$
and only the leading term is important at small $\varepsilon$. Clearly,
this leading term can be reproduced exactly in our $O(4)$ symmetric ansatz,
and our result is exact at small $\varepsilon$.  At very high energies
(small values of $\rho$) our field configuration tends to the $O(4)$
symmetric solution of the massless equations, Eq.(\ref{Khl}). So, at
$\varepsilon\gg 1$ it approaches the lower bound derived from this
 solution.  The alternative lower bound on $f(\varepsilon)$ can be easily
 read out from ref.\ \cite{Volb} and it is also shown in Fig.1.  This bound
has been obtained by direct analysis of diagrams. As one can see from
Fig.1, our bound is stronger than that of ref. \cite{Volb}.

\section{Conclusions}

To summarize, the tree multiparticle cross sections can be calculated using
a semiclassical method based on singular solutions to classical field
equations. At low kinetic energies $\varepsilon$ of particles in the final
state the corresponding solutions can be obtained analytically and the
leading exponent in the cross section can be found at least up to
$O(\varepsilon^2)$.  Instead of proceeding to higher orders in
$\varepsilon$, one can solve the field equations numerically for different
surfaces of singularity and after that apply the variational procedure at
arbitrary $\varepsilon$.  For the simplest form of the singularity, the
spherically symmetric one, the variational procedure over collective
coordinates is incomplete and provides a lower bound on the tree cross
section (the correct result at tree level should correspond to the true
maximum of the exponent). To improve this estimate, one should consider
more general field configurations. Beyond the tree level, the semiclassical
Landau-like methods are still applicable, but the variational procedure is
more complicated and corresponding solutions are known only in the simplest
case of zero energies.

The authors are indebted to
V.Rubakov for numerous helpful discussions. We would like to thank
M.Polyakov, P.Tinyakov and M.Voloshin for
valuable discussions. This work is supported in part by ISF grant \# MKT
300. The work of M.L.,  D.T.S., and S.T.\ is supported in part by
INTAS grant \# 94-2352. The work of F.B.\ is supported in part by Soros
fellowship for  students, and the work of M.L.\
and S.T.\ is supported in part
by Soros fellowship for graduate students.

\section*{Appendix }

The first two terms in the expansion (\ref{9}) have been obtained in
\cite{Son},
\begin{equation}
\varphi= \varphi_0+\varphi_1=
\sqrt{\frac{2}{\lambda}}
\left\{
\frac{1}{\sinh z}
+\frac{1}{2}\left(\partial^2\tau_0(\frac{\cosh z -\sinh z}{3}-
\frac{1}{\sinh z}+
z\frac{\cosh z}{\sinh^2 z}
)
+(\partial\tau_0)^2 z
\frac{\cosh z}{\sinh^2 z}\right)\right\}
\label{1'}
 \end{equation}
($z=\tau-\tau_0({\bf x})$). The function $\varphi_1$ satisfies the
following equation,
\begin{equation}
\partial_\tau^2\varphi_1-\varphi_1-3\lambda\varphi_0^2\varphi_1=
-\partial_i^2\varphi_0=-\partial_\tau^2\varphi_0(\partial_i\tau_0)^2+
\partial_\tau\varphi_0\partial_i^2\tau_0
\label{2'}
\end{equation}
The solution (\ref{1'}) can be found with the help of the Green function,
\[
\varphi=\varphi_0+\int^{z}d\sigma K(z,\sigma)(\partial_\sigma
\varphi_0 \partial^2\tau_0-\partial_\sigma^2\varphi_0(\partial\tau_0)^2)+
C_1(x)f_1(z)+C_2(x)f_2(z)
\]
where
\[
f_1(\tau)=-\partial_\tau\varphi_0=\sqrt{\frac{2}{\lambda}}\frac{\cosh(
z)}{\sinh^2(z)},
\,\,\,\,\,\,
f_2(\tau)=\sqrt{\frac{2}{\lambda}}\frac{\cosh(z)}
{\sinh^2(z)}\left(\frac{\sinh 2(z)}{4}-\frac{3(\tau-
\tau_0)}{2}+\tanh(z)\right)
\]
are the solutions of the homogeneous Eq.(\ref{2'}) and
\[K(z,\sigma)=\frac{\lambda}{2}(f_2(z)f_1(\sigma)-f_2(\sigma)
f_1(z))\]
is the Green function. The functions $C_1(x)$ and $C_2(x)$ can be found
from the conditions $\varphi(\tau\to \infty,x)\to0$ and $\varphi(0,0)\to
\infty$.

The next correction (we expect that it is of order of
$(\partial^4\tau_0)$) can be obtained in the same way from the following
equation,
\[
\partial_\tau^2\varphi_2-\varphi_2-3\lambda\varphi_0^2\varphi_2=
-\partial_i^2\varphi_1+3\lambda\varphi_1^2\varphi_0\]
and has the following form,
\[
\varphi_2=\sqrt{\frac{2}{\lambda}}\left\{
-\partial_i^2\partial_j^2\tau_0
\left({1\over 8}{1\over\sinh z} -{1\over 8} (\cosh z-\sinh z)-{1\over 12}z
(\cosh z-\sinh z) + {1\over 4}{z\over\sinh z}-{1\over 8}z
{\cosh z\over\sinh^2 z}
\right.\right.
\]
\[
-\left.{1\over 8}z^2{\cosh z\over\sinh^2 z}\right)
-\partial_j^2(\partial_i\tau_0)^2
\left({1\over 12}{1\over\sinh z}-{1\over 3}z{\cosh z\over\sinh^2 z}-{1\over
 12}z\cosh z +I(z)\right)
+2\partial_j\partial_i^2\tau_0\cdot\partial_j\tau_0
\left({1\over 12}{1\over \sinh z} \right.
\]
\[
\left.
-{1\over 3}z{\cosh z\over\sinh^2 z}
 -{1\over 4}{z\over\sinh z}+{1\over 4}z^2{\cosh z\over\sinh^2
z}-{1\over 12}z\sinh z+I(z)\right)
+2\partial_j[(\partial_i\tau_0)^2]\cdot\partial_j\tau_0
\left({1\over 8}{1\over\sinh z}-{1\over 8}z{\cosh z\over\sinh^2 z}
\right.
\]
\[
\left.-{1\over  24}(\cosh z-\sinh z)+{1\over 8}z^2{\cosh z\over\sinh^2 z}
\right)+\partial_i^2\tau_0\partial_j^2\tau_0
\left({1\over 12}z\cosh z -{1\over 4}z\sinh z+{1\over 4} (\cosh z-\sinh z) -
{3\over 8}{1\over\sinh z}
\right.
\]
\[
-{3\over 8}z {\cosh z\over\sinh^2 z}-{1\over 2}{z\over\sinh z}
\left.
 +{1\over 4}{z^2\over \sinh^3 z} +{1\over 8}{z^2\over\sinh
z}+{1\over 4}z^2 {\cosh z\over\sinh^2 z}+2I(z)\right)
+(\partial_i\tau_0)^2\partial_j^2\tau_0  \left({1\over 12}z
(\cosh z-\sinh z)\right.
\]
\[
\left. -{1\over 12}(\cosh z-\sinh z)+
{1\over 2}{z^2\over\sinh^3 z}+{1\over 4}{z^2\over\sinh z}
+ {1\over 4}{1\over\sinh z}
-{3\over 4}z{\cosh z\over\sinh^2 z}\right)
\]
\begin{equation}
\left.
-(\partial_i\tau_0)^2(\partial_j\tau_0)^2\left({3\over 8}z{\cosh
z\over \sinh^2 z}-{1\over 4}{z^2\over\sinh^3 z} -{1\over 8}{z^2\over\sinh
z}\right)\right\}
\label{2*'}
\end{equation}
where
\[
I(z)= {1\over 12}\ln(2\sinh z)\left(\sinh z+{3\over\sinh z}-3z{\cosh
z\over\sinh^2 z}\right)+
{1\over 4}{\cosh z\over\ \sinh^2 z}\int_0^z x \coth x dx .
\]

\end{document}